\documentclass{article}
\usepackage{times}
\usepackage{graphicx}
\begin{document}
\phantom{.}\vspace{0.4cm}
\large
Yamada Conference LVI, the Fourth International Symposium on Crystalline
Organic Metals, Superconductors and Ferromagnets \newline
ISCOM 2001 - Abstract Number B20Tue

\begin{center}
{\LARGE\bf 
Magneto-optics of organic crystalline field effect transistors
}
\end{center}

\begin{center}
{\Large A Ardavan$^{a,*}$, SJ Blundell$^a$ and J Singleton$^a$ }
\end{center}

\begin{flushleft}
$^a$ Clarendon Laboratory, Department of Physics, University of
Oxford, Parks Road, Oxford OX1 3PU, UK

\end{flushleft}

{\normalsize
$^*$ Tel; +44 1865 272364; Fax: +44 1865 272400; E-mail:
arzhang.ardavan@physics.ox.ac.uk 
}

\vspace{0.5cm}

\hrule

{\bf Abstract} \newline 

Organic molecular FETs provide an experimental framework for studying
the band-filling dependent properties of two-dimensional metals. Here,
we propose experiments to investigate the band-filling dependent
cyclotron resonance, and the effect of van Hove singularities.

{\bf Keywords:} Field-effect transistor, cyclotron resonance, van Hove
singularity. 

\vspace{0.5cm}
\hrule


In recent work, the group of Sch\"on, Kloc, Batlogg {\it et al}.\ have
taken a new approach to creating organic molecular metals, by gating
crystalline organic semiconductors in a field effect transistor (FET)
geometry~\cite{pentacene-fet-etc}, and effectively ``doping'' the
crystal by means of an electric field.  The resulting metals differ
from conventional organic molecular metals (where the ``chemical
doping'' arises from charge transfer between the cation and
anion~\cite{ishiguro}) in two important respects: The systems appear
to be exactly two-dimensional (i.e.\ the gate induced charge is
confined to the surface molecular layer) and the carrier density is
controllable over a wide range by adjusting the gate voltage. In
comparison to conventional FETs (e.g.\ Si-MOSFETs), the organic
crystals have a rather large unit cell. This means that a particular
induced carrier density corresponds to a substantially larger
band-filling; the band-filling dependent properties of the metal can
be investigated.  This approach has been very fruitful, yielding,
among other things, the integer and fractional quantum Hall effects in
a pentacene FET, superconductivity at 52~K in C$_{60}$, and a
tetracene laser~\cite{pentacene-fet-etc}.

\begin{figure}
\includegraphics*[width=15cm]{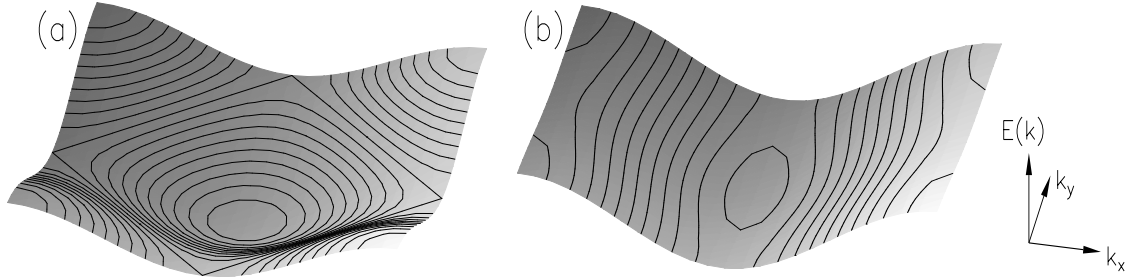}
\caption{The two dimensional tight-binding dispersion on a square
lattice for (a) isotropic hopping, $t_a=t_b$, (b) anisotropic hopping,
$t_a>t_b$.}
\label{dispersions}
\end{figure}

The single-particle dispersion in systems of this kind is well
described by a tight-binding model approach~\cite{a+m}. For a
rectangular lattice with lattice constants $a$ and $b$ and transfer
integrals $t_a$ and $t_b$ in the $x$- and
$y$-directions respectively, the dispersion is 
$ {\mathcal E}({\bf k})=-2t_a\cos(k_x a)-2t_b\cos(k_y b) $ .
This is shown in Figure~\ref{dispersions} for a square lattice
($a=b$), for the isotropic case (a) $t_a = t_b$ and the anisotropic
case (b) $t_a > t_b$. The solid lines are energy contours,
representing the possible Fermi surfaces that can be induced by
gating. 
For the isotropic dispersion, the possible Fermi surfaces are all
closed, changing from electron-pockets to hole-pockets as the
band-filling is increased. In the anisotropic case, however, closed
electron or hole pockets only exist when the Fermi energy is within
$4t_b$ of the bottom or top of the band repectively. For intermediate
values, the Fermi surface is open. For most real (lower symmetry)
two-dimensional dispersions, the fact that the dispersion is periodic
leads to the expectation that both open and closed Fermi surfaces can
exist.


In the semiclassical picture, a magnetic field applied perpendicular
to the two-dimensional metal causes the charge carriers at the Fermi
surface to orbit the Fermi surface under the influence of the Lorentz
force~\cite{a+m}. The components of the carriers' real-space velocity
(given by the $k$-space gradient of the dispersion, ${\bf
v}=\hbar^{-1}\nabla_k {\cal E}$) oscillate at the cyclotron frequency,
and this time dependence determines the frequency-dependent
conductivity, $\sigma(\omega)$, as described by Chambers' formulation
of the Boltzmann transport
equation~\cite{sjb-aa,mck-freq-dep}. Each harmonic component of
the time evolution of the real-space velocity, with frequency
$\omega_{n}$, generates a cyclotron resonance in $\sigma(\omega)$ when
$\omega=\omega_n$.
\begin{figure}
\includegraphics[width=15cm]{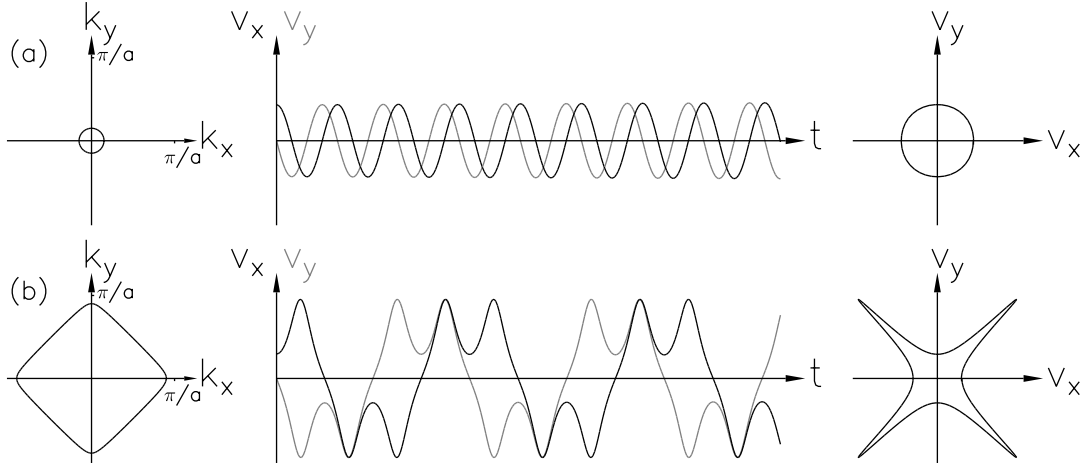}
\caption{The real-space velocity evolution calculated for two values
of the bandfilling in the dispersion shown in
Figure~\ref{dispersions}(a). Left: the Fermi surface. Middle: the time
dependence of $v_x$ and $v_y$. Right: the locus of points $(v_x,v_y)$.}
\label{2dsim}
\end{figure}

Figure~\ref{2dsim} shows the time evolution of the components of the
real-space velocity of a carrier in the dispersion shown in
Figure~\ref{dispersions}(a), for two values of the
band-filling. Figure~\ref{2dsim}(a) shows the case for which the Fermi
energy is close to the bottom of the band. The dispersion is close to
parabolic and the real-space velocity evolves sinusoidally with
time, leading to a single cyclotron resonance. 
Figure~\ref{2dsim}(b) shows the case for which the band
is almost half-full. The magnitude of the Fermi velocity varies
substantially around the Fermi surface. Consequently the time
evolution of the real-space velocity components is no longer
sinusoidal, containing rich odd-harmonic content~\cite{sjb-aa}. This
leads to the presence of harmonics of the fundamental cyclotron
resonance at frequencies $\omega=n\omega_c$, with $n$
odd~\cite{sjb-aa}. Note that the cyclotron period is substantially
increased over case (a).

\begin{figure}
\includegraphics[width=15cm]{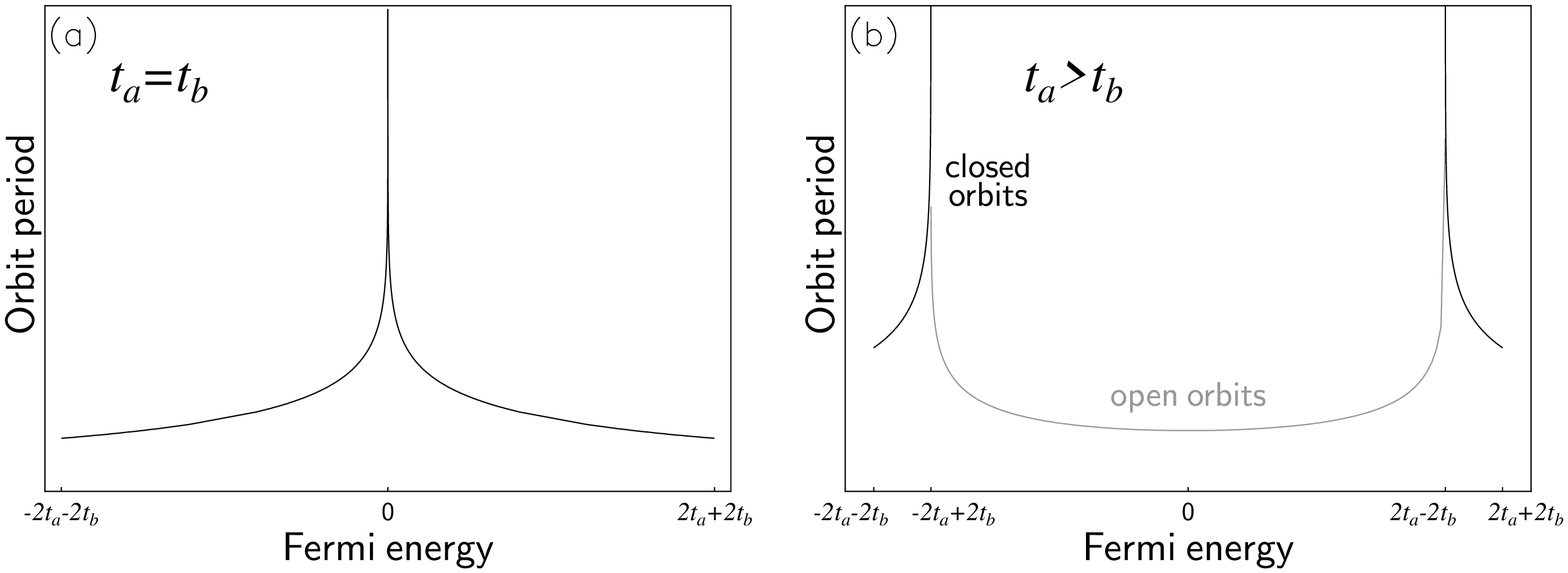}
\caption{The cyclotron orbit period as a function of
band-filling calculated for the dispersions in Figure~\ref{dispersions}.}
\label{periods}
\end{figure}

Figure~\ref{periods} shows the dependence of the cyclotron period on
band-filling for the cases (a) $t_a=t_b$ and (b) $t_a > t_b$.
In each case the cyclotron period is strongly dependent on the
band-filling. When $t_a \neq t_b$, open orbit cylotron resonance
(Fermi surface traversal resonance)~\cite{1dcr} is
possible over a range of band-filling (the period is shown in grey in
Figure~\ref{periods}(b)). The period diverges when the Fermi energy
coincides with energies for which $\nabla_k {\cal E} \rightarrow 0$
(van Hove singularities~\cite{a+m}). This divergence arises because
for these values of the Fermi energy, there exist points on the Fermi
surface for which the magnitude of the Fermi velocity, and hence the
Lorentz force, vanish. 

In a quantum mechanical picture, ${\bf k}$, which labels the crystal
eigenstates, is in general no longer a good quantum number in the
presence of a magnetic field. The new eigenstates, Landau levels, are
delocalized in $k$-space.  However, the semiclassical picture suggests
that the $k$-states close to a van Hove singularity remain stationary
even when a magnetic field is applied. The state of the
two-dimensional system in magnetic field therefore appears to change
qualitatively when the Fermi energy coincides with a van Hove
singularity, and the consequent effect on the macroscopic properties
remains to be experimentally tested. To our knowledge, organic
crystalline FETs provide the only experimental system in which this
condition can be approached controllably.

\end{document}